# Integration, Verification and Layout of a Complex Multimedia SOC

Chien-Liang Chen
Global UniChip Corp.
Hsin-Chu Science Park
Hsin-Chu, Taiwan 300
+886-3-564-6600

clchen@globalunichip.com

Jiing-Yuan Lin
Global UniChip Corp.
Hsin-Chu Science Park
Hsin-Chu, Taiwan 300
+886-3-564-6600

jylin@globalunichip.com

Youn-Long Lin
National Tsing Hua University
Dept CS, NTHU
Hsin-Chu, Taiwan 300
+886-3-573-1072

ylin@cs.nthu.edu.tw

## ABSTRACT
We present our experience of designing a single-chip controller for advanced digital still camera from specification all the way to mass production. The process involves collaboration with camera system designer, IP vendors, EDA vendors, silicon wafer foundry, package & testing houses, and camera maker. We also co-work with academic research groups to develop a JPEG codec IP and memory BIST and SOC testing methodology. In this presentation, we cover the problems encountered, our solutions, and lessons learned.

## 1. INTRODUCTION
Ever increasing computational demand from the application side and very deep submicron semiconductor processing from the technology side together make system-on-chip (SOC) reality and necessary. Makers of such electronics systems as PDA, cellular phone handsets, digital still camera, etc, need ASIC solutions in order to differentiate themselves from the competition and to cut cost. On the other hand, semiconductor foundry has to expand its service scope from wafer manufacturing to mask tooling, cell & I/O library, memory compiler, and up to IP such as PLL, DAC, and ADC. Therefore, there is a need to bridge the gap between system houses and wafer foundry. We call such company SOC design service provider.

An SOC design service provider takes as its inputs from the system house a specification or partially-designed prototype and delivers to its customer GDSII ready for manufacturing. Close collaboration is needed among all parties in order to successfully bring a competitive product to the market in time.

## 2. A DIGITAL STILL CAMERA SOC
Our objective was to design a single chip controller for 2-million-pixel and 3-million-pixel grade DSC for mass production of 3.5 million units in a year. In order to satisfy desired functionality at low cost, the SOC was specified to include the following IPs: a hybrid RISC/DSP processor, a hardwired JPEG encoding and decoding engine, a USB 1.1. device/mini-host controller with TxRx PHY, an SD/MMC flash card host interface, an SDRAM controller, an LCD Interface, an NTSC/PAL TV encoder, a 10-bit video DAC, an 8-bit LCD DAC, and two PLLs. The IPs come from multiple sources. Each of them posts different challenge to the project team.

The hybrid RISC/DSP was not an IP at all. Actually it was a stand alone processor chip used in the previous generations of cameras. To meet high speed requirement (133MHz @ 0.25um), we have to make it a hard core before integration with others. To integrate it into the SOC, we have to collaborate with the original vendor to create synthesis, simulation and test models in addition to hardening the processor into a high-speed hard macro.

The USB1.1 controller and the SD card controller are supplied by a third party vendor. They are in VHDL RTL instead of more locally popular Verilog. Therefore, mixed-language simulation environment has been set up.

To meet processing speed requirement of 3M pixels @ 0.1Sec and long battery life, the JPEG codec function has been implemented in a hardware accelerator. We collaborate with a university research laboratory, which will present the IP in more detail in a companion paper [1] of this session. The effort we spent was in bridging the gap between university prototype and industrial strength design.

After all IP models are made ready, whole system integration and verification is an even bigger challenge. We encountered the problem of in-consistent and in-sufficient test benches. Therefore, developing test bench as the project goes is very important. Tight co-operation among all parties and precise project management are essential to complete the integration and verification of the system.

## 3. SILICON IMPLEMENTATION
The DSC controller consists of 240K gates excluding memory macros. After whole system verification with hybrid emulation/simulation, it was implemented in TSMC 0.25um 1P5M CMOS process and packed in TFBGA256 package. It took three months for a team of six engineers to complete the Netlist-to-GDSII service. During the course, there are many changes to the spec and netlist. Therefore, the implementation team has to be flexible and adaptive to changes.

There are 30 embedded memory macros in the controller. We use an in-house memory BIST circuit generator to insert one common BIST controller, multiple sequencers, and 30 pattern generators. The MBIST is from collaboration between us and a university research laboratory, which will also describe the technology in detail in another companion paper [2]. After scan insertion, the fault coverage was 93%.

The physical design of the chip was done with timing-driven placement and routing, physical synthesis, formal verification and STA QoR check.

During chip implementation, we encountered several problems:

- During the course, there are 3 spec changes involving re-synthesis and FF modification, 10 netlist changes involving ECO of combinational logic part, 3 ECO changes to fix setup/hold time violation, and 13 versions of pin assignments.
- There existed inconsistency between simulators/versions among customer, IP vendors and us. The customer used PC-based Verilog/ModelSim while we used NC-Verilog. This lead to extra twist during ASIC sign-off.
- IP quality is less than ideal. We have to clean up many DRC/LVS violation in the database provided by the IP vendors.
- The USB IP was delivered in FPGA-targeted RTL. No robust synthesis script was available and the first RTL level simulation was failed. We have to co-work with the IP vendor over 10 versions of RTL code modification or synthesis constraint updates.
- Because there is no automation tool available, we manually performed many version of pin assignments to reduce the number of substrate layers from four to two resulting in packaging cost saving.

During mass production, manufacturing test uncovered that the yield killer (5% loss) was in the insufficient driving strength of an output buffer in the CPU. The chip also went through reliability test including ESD performance test, temperature cycle test, high/low temperature storage test and humidity/temperature test.

The mass production yield was enhanced from 82.7% initially to very close to foundry's yield model of 93.4% over a period of 8 months. Our measures included optimizing probe card overdrive spec, optimizing power relay waiting time, and retargeting $I_{sat}$ and $V_{th}$ by optimizing poly CD in the foundry according to results from corner lot splitting. We also corrected the insufficient driving strength problem by means of metal changes to utilize the spare cells.

We have been requested to perform failure analysis on 20 returned chips that have pins shorted to GND. After checking substrate delaminating and popped-corner using scanning acoustics tomography, we found no abnormality. Finally, by sinking 400mA of current to the corresponding pin of a good chip we concluded that the failure was due to a system board bug.

## 4. CONCLUSION

We went on to produce over three millions of the chip over 18 months. Our system customer was able take about 8% of world-wide market share during that period. We have also migrated the chip from 0.25um process to 0.18um one achieving 20% saving in die cost.

The project has successfully demonstrated that it is feasible to bridge the gap between the need of an electronics system house without IC design capability and the production capacity of a semiconductor foundry with an SOC design service provider. We have been able to leverage the experience gained and lesson learned to serve more customers and more projects such as DVD player, cellular phone set, digital TV, photo display, etc.

The project experience told us that it is quite risky to employ third party IP in a complex SOC project, especially, when the IP has not been proven in the identical design environment.

As both applications and technology become more advanced, we have expanded our IP portfolio to include MPEG4, USB2.0 and OTG, SerDes I/O and embedded Flash memory. We have also enhanced our EDA flow to be able to handle dozens of multi-million gate design at 0.13um process and below every year.

Current complex SOC projects require silicon implementation flow including virtual prototyping, signal integrity check (crosstalk, electron-migration, dynamic IR drop, de-coupling cell insertion), design for manufacturability (intra-die process variation modeling, double via, dummy metal insertion), STA sign-off with in-die variation analysis, hierarchical DFT and physical implementation, low power solution (multi $V_t/V_{DD}$ cell library, gated clock, power down isolation) and flip-chip solution.